# Connection of converters to a low and medium power DC network using an inductor circuit

F. Pérez-Hidalgo, J.R. Heredia-Larrubia, A. Ruiz-González, M. Meco-Gutiérrez and F. Vargas-Merino

*Abstract*—This letter describes an alternative for the connection of power converters to a direct current network without the installation of a capacitor in the DC-Link. The circuit allows the connection of converters through a coil and avoids short-circuit currents with different instantaneous values of voltage output. A description of the calculation and the choice of components together with a real implemented example in a DC network within Smart City project (*Endes Utility*) is presented.

## I. INTRODUCTION

Renewable energy sources are increasingly used for various reasons [1]. The most important contribution to the energy mix is from wind power and photovoltaic. In the last years, it has also been seen how the installation of distributed generation, highly promoted in the country, has been increased exponentially in the medium and low voltage networks. In the majority of applications, connections are made to the network using power converters. It is noteworthy that these forms of energy are increasingly popular and are used in self generation, or to generate energy for input to the grid from a residential installation. In the latter case, power-flow management is required to optimize the energy system. This uses a control that governs the connection and disconnection of energy sources. Currently these ideas are assembled under the topic "smart grid" [2]. *Endesa* creates the SmartCity Málaga project. SmartCity Málaga is a project developed on the real Power Grid of *Endesa*, in the city of Málaga, Spain. Therefore, it is a Project based on the optimization and modernization of the current Distribution Network [3].

In this pilot project customers and Distribution Company collaborate in the achievement of the energetic challenges, increasing the use of renewable energy sources, moving generation closer to consumption and pledging for a rational and efficient consumption, all combined with storage facilities, giving support to the grid in critical moments. In these facilities, DC links are increasingly used as generation comes from a DC source (i.e. photovoltaic) and/or from a rectified AC supply (i.e. wind) [4][5] and battery. All connections are made via power converters, rectifiers or DC/DC converters and subsequently connected to the AC mains through power Inverters [6]. Usually in the DC-Link a capacitor is used for various reasons. But increasingly we see the possibility of DC networks where currents converge generation from different energy sources, both low voltage and medium voltage (MVDC). In the DC-link it is difficult to calculate the real value of the capacitors when taking into account that the network can be extended with the connection of new generation sources. In addition, the capacitors are expensive. One of the problems with the connection of power converters to a DC network is that although the average voltage values are equal, the instantaneous values need not be matched, causing short circuits if they are connected directly to the DC-Link. This is usually avoided with the use of capacitors. The purpose of this article is to present a circuit design based on an inductor to connect the converters to the DC-Link that has no need for the use of a condenser.

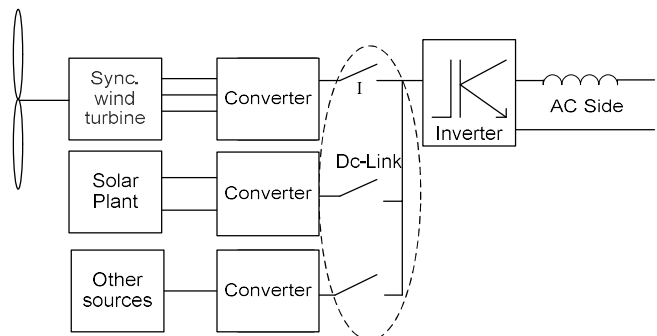

Fig. 1 *DC-Link with different sources of generation*

## II. CIRCUIT DESCRIPTION.

A diagram of a DC-Link where different generation sources may converge is shown in Fig.1. As seen in the figure, each converter will have a switch to connect it through to the DC-Link. With this scheme the connection may produce a short circuit if there is no capacitor in the DC-Link. This can be resolved by connecting the converters using an inductor for each one of the converters. This solution is more flexible because it allows easy expansion and is also more economical because the sizing is carried out for each of the converters. The drawback of this solution is the disconnection.

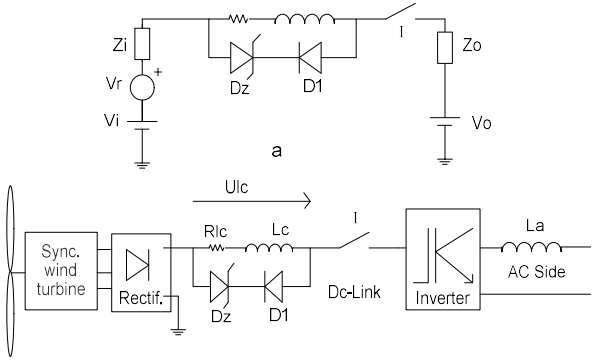

**Fig. 2** *Connection circuit.*
*(a) Equivalent circuit*
*(b) A wind generator connected to the DC-Link circuit*

The converter cannot be disconnected abruptly if it is supplying current to the DC-Link as this would cause surges and arcs in switch I. To overcome this drawback, it will have a current-limiting circuit of the inductor by means of two diodes, one of them a zener diode. The diagram of this solution is shown in Fig. 2-a. The inductor electrical parameters are first calculated. This proceeds from the basis that the converter injects to the network one continuous current plus a harmonic content. The equivalence of this behaviour is given by the superposition of a DC voltage source $V_i$ in series with a source representing the harmonic content or the distortion, $V_r$. The Thevenin equivalent impedance of the converter is $Z_i$ (Fig. 2-a). Similarly, the circuit of the DC network is given by the equivalent $V_o$ in series with an impedance $Z_o$. The inductor $L_c$ connects between the two equivalent circuits. This coil will also be an effective reducer of harmonics emitted by the power converter.

This inductor is aimed at avoiding short circuits due to different instantaneous values between the converters discussed above, but it is also exploited for use as a low-pass filter and to reduce harmonics. The case of $Z_o$ and $Z_i$ tending towards zero, would be the worst case for the calculation of the inductor and the source impedances will be avoided. As expected, and according to the technical literature [7], to reduce the harmonics emitted to the network by the power inverters, an inductor of 3% ($L_a$) relative to the baseline values of power and voltage of the generating source ($Z_{base}$), is typically used on the AC side, thereby *ITDH* is reduced to approximately 50% (Fig. 2-b). To obtain a similar percentage reduction of the *ITDH* in the DC side an inductor $L_c$ equal to the value of $L_a$ is used but multiplied by 1.7 times [7]. These calculations are easily resolved if all the parameters of the sources are known, line voltage, $V_{ll}$ in kV and apparent power, $S$ in kVA.

$$Z_{base} = \frac{(V_{ll}^2 \cdot 1000)}{S} \quad (1)$$

$$\frac{3\%}{100} = \frac{X_{la}}{Z_{base}} \quad ; \quad X_{la} = L_a \cdot 2\pi \cdot 50; \quad L_c = 1.7 * L_a \quad (2)$$

During the disconnection, energy stored in the coil is discharged through the circuit formed by the coil and the two diodes. The voltage across the coil $U_{lc}$ is given by the expression:

$$U_{lc} = L_c \frac{di_{lc}}{t_d} = U_{D1} + U_{Dz} \quad (3)$$

$I_{lc}$ is the current that circulates through the inductor and $U_{D1} + U_{Dz}$ is the voltage drop in the diode.

On the other hand, is knon that in the circuit the maximum voltage that the inductor can withstand will depend on the maximum DC current $I_{lcmax}$ and on the currents of the harmonic content and the maximum power $P_{lcmax}$.

$$U_{lcmax} = \frac{P_{lcmax}}{I_{lc}max + \sum_{i=1}^{n} i_{ci}} = U_{Dz} + U_{D1} \quad (4)$$

*Application*. This was mounted in a domestic DC network within a smart grid project (known as Smart City) in Málaga (Spain), owned by the *ENDESA Utility* with different DC generation sources (wind, photovoltaic and battery). Below is an example of the calculation of the elements of the circuit in the case of a synchronous wind generator of 0.8 $K_{VA}$ connected to a network of 48V DC through a 6-pulse power rectifier (Figure 2-b). The reactance $L_a$ for power and voltage values was calculated at 3%. The $Z_{base}$ is given by equation 1.

$$Z_{base} = \frac{(0.048^2)1000}{0.8} \quad (5)$$

The DC network is connected to a 50 Hz AC network through an inverter-transformer. For the calculation a 1:1 relationship of the inverter-transformer was assumed for no change of basis between the AC and the DC sides, as calculating the value $L_c$, is of primary interest. The reactance $X_a, L_a$ and $L_c$ are calculated using expressions 2.

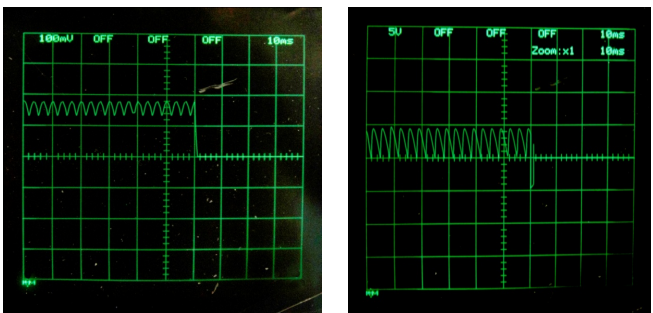

**Fig. 3**: *Experimental results of the disconnection of the inducer.*
*(a) Current 10ms/div; 10mV=1A/div (b) Voltage 10ms/div; 5V/div*

$$L_a = 270\ uH;\quad L_c = 459\ uH \tag{6}$$

A commercial inductor of 500uH that supports a maximum current of 20A was used, higher than the generator can deliver. The value of $R_{lc}$ for this inductor is 0.2 Ω. The maximum power $P_{lcmax}$ which the inductor $L_c$ can dissipate is 80W.

Knowing the maximum DC values for the generator and the currents of the first 50 harmonics of the ripple of the 6-pulse rectifier, a $U_{lcmax}$ of 4.7V is obtained with expression 4. This value helps with the choice of the diodes.

For this case, a 1N5335B Zener diode with a Zener voltage $U_z$ of 3.9 V was chosen, to withstand an $Izsm$=17.6 A for a maximum time of 8.3 ms. The SBR20A200CTB diode was also chosen with 20 A and 0.5 V of forward drop and that supports 100V in inverse dynamic resistance of $R_d$=1.34 mΩ.

The inductor current during disconnection is shown in Fig.3-a. In Fig.3-b the voltage in the coil is in parallel with the diodes. It can be seen that the maximum supported voltage does not go beyond 5 V value.

### III. Conclusion

This letter presents the description and calculation of a circuit for connection and disconnection of power converters to a DC-Link without capacitors. The displayed solution is more flexible, economical and expandable than the use of capacitors for DC connection of different sources of renewable energy. The feasibility of the proposed circuit has been validated in a real case proving the benefits described above.